\documentclass[12pt,a4paper]{article}
\usepackage{amssymb,amsmath}
\usepackage[T2A]{fontenc}
\usepackage[cp866]{inputenc}
\usepackage[russian,english]{babel}
\usepackage{graphicx}
\newcommand{\risheight}{10cm}
\begin{document}
\begin{center}
{\Large Why is ${}^{208}_{82}{\rm Pb}$ the heaviest stable nuclide?}\\[3mm]
{B.~P.~Kosyakov${}^{a}$, E.~Yu.~Popov${}^a$, and  M. A. Vronski{\u\i}${}^{a,b}$}\\[3mm]
{{\small ${}^a$Russian Federal Nuclear Center--VNIIEF, 
Sarov, 607188 Nizhni{\u\i} Novgorod Region, Russia;\\
${}^b$Sarov Institute of Physics {\&} Technology, Sarov, 607190 Nizhni{\u\i} Novgorod Region, Russia}\\} 
\end{center}
\begin{abstract}
\noindent
{In an effort to understand nuclei in terms of quarks we develop an effective theory to low-energy 
quantum chromodynamics in which a single quark contained in a nucleus is driven by a mean field due 
to other constituents of the nucleus. 
We analyze the reason why the number of $d$ quarks in light stable nuclei is much the same as that 
of $u$ quarks, while for heavier nuclei beginning with  ${\rm {}^{40}_{20}Ca}$,  the number of $d$ quarks is greater than the number of  $u$ quarks.
To account for the finiteness of the periodic table, we invoke a version of gauge/gravity duality between the dynamical affair in stable nuclei and that in extremal black holes. 
With the assumption that the end of stability for heavy nuclei is dual to the occurrence of a naked singularity, we find that the maximal number of protons in stable nuclei is 
$Z_{\max}^{\rm H}\approx 82$.}
\end{abstract}


\section{Introduction}
\label
{Introduction}
In the early 1930s, the view of a nucleus as a bound system of neutrons and protons \cite{Iwanenko}, \cite{Heisenberg}, held together by meson exchanges \cite{Yukawa}, became well-accepted.
This view, with several innovations, such as spontaneously broken chiral symmetry, effective Lagrangians, and derivative expansions~\cite{Weinberg1990}, remains a pillar of modern nuclear physics~\cite{Epelbaum}, \cite{Machleidt}.
With the advent of quantum chromodynamics (QCD) a serious effort was mounted to describe nuclei in terms of quarks.
The reason for this is twofold.
Firstly, we are still far from understanding the structure of the periodic table. 
Why do stable light nuclei contain equal parts of protons and neutrons? 
Why are all nuclei  heavier than  ${\rm {}^{40}_{20}Ca}$ neutron-rich?
Why is every nucleus with the number of protons $Z$ above $Z=82$ unstable~\footnote{The term `stability' is taken here to mean `absolute stability'.
A nuclide with very long lifetime $T$, say $T\sim 10^{20}$ years, is regarded as unstable.
We think of free protons as stable particles.}?
Secondly, there is direct evidence that a quark confined to a nucleus is not forced to live in a triple room.
Indeed, a free neutron, when combined with a proton to form a deuteron, loses the responsibility for the fate of its own quarks. 
This is clear from the fact that the $d$-quark lifetime relative to $\beta$-decay increases from $T\approx 15$ minutes to $T=\infty$.

The simplest way for introducing quarks into nuclear physics is to conceive of a nucleus with mass number ${A}$ as a bound system of ${\cal N}=3{A}$ quarks enclosed in a `bag' of size $R$ ~\cite{Petry}.
However,  the stability of the bag stipulates \cite{Close} that ${\cal N}$ and $R$ must be related by $R\sim{\cal N}^{1/4}$, contrary to the firmly established \cite{Angeli} phenomenological relation  
\begin{equation}
{R}={R_0}\, A^{1/3}\,,
\label
{R-nucleus}
\end{equation}                                           
and the discord is particularly noticeable for heavy nuclei.
Furthermore, the magnetic moments of such bags differ from the experimentally measured magnetic moments of nuclei \cite{Arima}, \cite{Talmi}.
An effort to account for the properties of nuclei by eliminating gluon degrees of freedom was reasonably successful~\cite{Maltman} but never progressed beyond small nuclei.

Another way of looking at stable nuclei \cite{KPV-1}, \cite{KPV-2} is to integrate out irrelevant degrees of freedom,  and consider only field variables $\Psi$ of a {single quark} contained in a studied nucleus. 
This quark, driven by a mean field generated by all other constituents of the nucleus, is to be responsible for the static properties of the nucleus. 
The dynamics of the quark is assumed to be encoded in the action
\begin{eqnarray}
{\cal S}
=\int d^4x\left\{{\Psi}^\dagger\left[\gamma^\alpha\left(i\partial_\alpha+g_VA_\alpha\right)-m\right]\Psi 
+g_S{\Psi}^\dagger\Psi\,\Phi\right\},
\label
{QCD-Lagrangian}
\end{eqnarray}                         
where  $A_\alpha=(A_0,-{\bf A})$ and $\Phi$ are respectively the Lorentz vector and scalar potentials of the mean field, and $g_V$ and $g_S$ their associated couplings.
The form of the action suggests that the color charge of the quark is in fact screened by a color-polarizable medium of nuclei, and hence the effective theory describes a color singlet entity, a fermionic quasiparticle bearing some resemblance to a polaron in condensed matter physics, rather than a quark appearing in the fundamental QCD Lagrangian. 
With this reservation, we still prefer to use the term `quark' for this carrier of the flavor quantum number.

We augment the dynamical description by the addition of the so-called `pseudospin symmetry condition' \cite{Ginocchio}, \cite{Liang}.
The purpose of this condition is twofold: (i) to convert the current quark mass into the constituent quark mass through a shift of mass, and (ii) to balance scalar attraction and vector repulsion of the mean field for confining the quark to nuclei \cite{KPV-3}.
The findings of these studies are briefly outlined in Sec.~\ref{systems of quarks}.

The use of `gauge/gravity duality', aka `correspondence between a theory of quantum gravity in anti-de Sitter space and conformal field theory in Minkowski space', `AdS/CFT', and `holography' \cite{Maldacena}--\cite{Klebanov} (for a full coverage of ideas and methods of gauge/gravity duality see \cite{Ammon}, \cite{Nastase}), opens a new avenue of attack on the problem.
Loosely speaking, the gauge/gravity duality is a doctrine which states that a good part of subnuclear physics in 4-dimensional Minkowski spacetime ${\mathbb R}_{1,3}$ is modelled on physics of black holes (BHs) and similar black objects~\footnote{A variety of such objects (black rings, black branes, etc.) has been detailed in \cite{Emparan}, \cite{Horowitz}.} in 5-dimensional anti-de Sitter space,  ${\rm AdS}_5$, whose boundary is just this ${\mathbb R}_{1,3}$. 

The mainstream develops the idea that a BH in ${\rm AdS}_5$ is holographically mapped onto a quark-gluon plasma lump in ${\mathbb R}_{1,3}$ \cite{Herzog}.
In an alternative approach, Dp-branes are mapped onto subnuclear entities in the confinement phase \cite{Witten-98}, \cite{Sakai}.
As to the present context, suitable objects here are extremal BHs in ${\rm AdS}_5$ which are mapped onto  microscopic {stable} systems in  ${\mathbb R}_{1,3}$, say {stable nuclei}
 \cite{KPV-3}.
This selection of dual objects substantiates the {pseudospin symmetry condition} \cite{KPV-3}.

Why is our concern with extremal BHs?
Recall that extremal BHs are free of Hawking evaporation.
Therefore, extremal  BHs and stable nuclei share a common trait, that of defying spontaneous ejection of their constituents.
The question arises of whether an ordinary BH amenable to Hawking evaporation can be dual to an unstable nucleus? 
The answer is no. 
Firstly, nuclear reactions are  {reversible}, whereas Hawking evaporation is irreversible. 
Secondly, quantum mechanical systems of the same species are identical and indiscernible.
It is expected that their gravitational duals exhibit similar properties.
Let two Lorentz frames, ${\cal F}$ and ${\cal F}'$, meet at some instant.
Assume that the ${\cal F}$ carries a Schwarzschild BH whose mass at this instant equals that of another Schwarzschild  BH attached to the ${\cal F}'$.
Are their masses always equal?
The rate of evaporation, as measured on the proper time, is common for both BHs.
Therefore, at a later time, simultaneous measurements of masses of the BHs attached to the 
${\cal F}$ and ${\cal F}'$ will give {different} results.  
The relativistic effect of time dilation keeps evaporating BHs from being regarded as 
identical entities.

The holographic mapping of bulk Dirac fermions in near extremal BHs was also analyzed in various contexts, (see, e.~g., \cite{Liu}--\cite{Mamo} and references therein), and the obtained results hold much promise for future gauge/gravity studies. 

In the present paper, we address the question of quark content in stable nuclei. 
It will be seen in Sect.~\ref{systems of quarks} that the idea of `quantum minimalism'~\footnote{Manifestations of the quantum minimalism in microscopic realm are numerous. 
For example, if the center-of-mass energy in a scattering process is of order of the Planck energy $E_{\rm Pl}\approx 10^{19}$ GeV but the momentum transfer is small compared to $E_{\rm Pl}$, all physical modes of the gravitational field are classical, except for two longitudinal modes \cite{Verlinde}. 
On the other hand, to develop an effective theory to low-energy QCD one may turn to the SU$({\cal N})$ Yang--Mills theory in the 't Hooft limit $g^2_{\rm YM}{\cal N}\to\infty$.
Quantum fluctuations disappear in this limit \cite{Witten79}, namely ``the measure in function space becomes concentrated on a single orbit of the gauge group'', and ``the probability of finding any gauge invariant quantity away from its expectation value goes to zero as ${\cal N}$ goes to infinity'' \cite{Coleman}.    
Therefore, the 't Hooft limit signifies that a system of strongly coupled quarks is governed by a semiclassical, feeble quantum dynamics.
We restrict the application of quantum minimalism concept to nuclear physics and  gauge/gravity duality.} furnishes insight into the problem of nuclear stability.
We argue that a sequence of stable nuclei of increasing $Z$ cannot be extended above some $Z_{\max}$ because this would contravene the line of demarcation between the quantum and classical regimes of evolution.
Section~\ref{Holographic} is devoted to calculation of $Z_{\max}$ on the assumption that $Z_{\max}$ is attained with the occurrence of a naked singularity in ${\rm AdS}_5$. 
The rationale behind this maneuver is the statement \cite{K-2008} that if the classical and quantum regimes of evolution coexist, then their interface is feasible at the event horizons of BHs. 
The disappearance of the event horizon and outcrop of a naked singularity imply that the classical-quantum arrangement is violated not only in ${\rm AdS}_5$ but---through the gauge/gravity duality---also in  ${\mathbb R}_{1,3}$.

Natural units, $\hbar=1, c=1$, are adopted throughout the text.
We use the conversion formula $m_p=0.938\,{\rm GeV}=\left(0.210\,{\rm fm}\right)^{-1}$.

\section{Nuclei as bound systems of quarks}
\label
{systems of quarks}
Is it possible to deduce the rule that a stable light nucleus is half-filled with protons?
In fact, among stable isotopes of a given light element of the periodic table there is always that specified by $Z=N$, except for beryllium represented by a single stable nuclide, ${\rm {}^{9}_{4}Be}$, with $N/Z=1.25$. 
We take, as the starting point, the semiempirical Weizs\"acker relation for the binding energy per nucleon,
\begin{equation}
\frac{B}{A}=\alpha-\beta\,A^{-1/3}-\gamma\, \frac{(Z-N)^2}{A^2}-\delta\,Z^2 A^{-4/3}\,.
\label
{weizsacker}
\end{equation}                                           
The coefficient of the next to the last term, associated with the repulsive effect due to the Pauli exclusion principle, $\gamma=23.3$ MeV, is $32$ times greater than $\delta=0.72$ MeV, the coefficient of the last term, associated with the Coulomb repulsion.
Therefore, the degeneracy pressure plays the defining role in the force balance, as opposed to the Coulomb interactions. 

Let us conceive of a nucleus containing $Z$ protons and $N=A-Z$ neutrons as a mixture of two Fermi gases.
The degeneracy pressure $P$ in this system is proportional to
\begin{equation}
{{Z}^{5/3}}\,{m_p}^{-1}
+
{\left(A-Z\right)^{5/3}}\,{m_n}^{-1}\,,
\label
{deg_pressure}
\end{equation}                                           
where $m_p$ and $m_n$ are the respective masses of protons and neutrons.
To deduce the desired rule, we note that $P$ varies in direct proportion to the energy density 
${\cal E}$, and require that ${\cal E}$ be minimal.
The requirement is obeyed by $Z=A\,m^{3/2}_p\left(m^{3/2}_n+m^{3/2}_p\right)^{-1}$.
Since $m_p\approx m_n$, this results in $Z\approx 0.5 A$.
However, this expedient is unsuited for a deuteron, because it is a system composed of a proton and 
a neutron, and there is nothing to be affected by the degeneracy pressure.

The above argument, reiterated in terms of $u$ and $d$ quarks available in the nucleus, shows that the 
number of $u$ quarks, $n_u$, which minimizes ${\cal E}$ is 
\begin{equation}
n_u=\left(n_u+n_d\right) m^{3/2}_u\left(m^{3/2}_u+m^{3/2}_d\right)^{-1}.
\label
{n-u-m-d-n-tot-over-m-tot}
\end{equation}                                           
If it is granted that $m_u\approx m_d$ (as it is for the constituent quark masses), then 
(\ref{n-u-m-d-n-tot-over-m-tot}) becomes 
\begin{equation}
n_u\approx n_d\,.
\label
{n-approx-n}
\end{equation}                                           
The rule (\ref{n-approx-n}) holds for a deuteron containing $n_u=3$ and $n_d=3$, but is no longer valid for a triton involving $n_u=4$ and $n_d=5$, and all the more for a dineutron, because it has $n_u=2$ and $n_d=4$.
The quark content of ${\rm {}^{9}_{4}Be}$ might be fitted reasonably well with (\ref{n-approx-n}) when one takes into account that $n_u=13$ is not too far removed from $n_d=14$ because $n_d/n_u\approx 1.08$.
Good agreement with the actual quark content of stable light nuclei suggests that the treatment of nuclei in terms of $u$ and $d$ quarks is realistic.

\begin{figure}[htb]
\centerline{\includegraphics[height=\risheight]{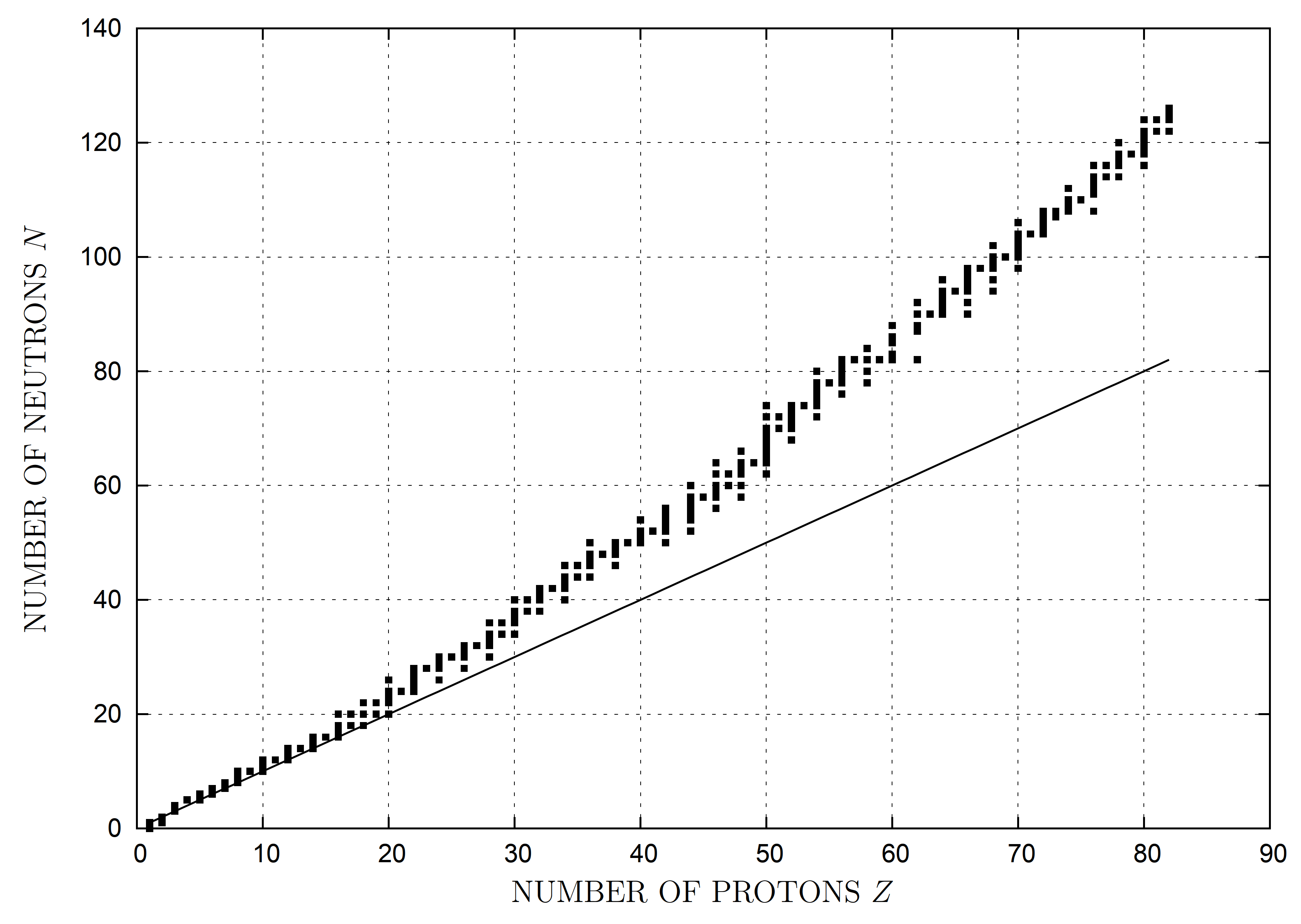}}
\caption{Plot of stable nuclides composed of $Z$ protons and  $N$ neutrons}
\label{stable}
\end{figure}

For $Z>20$, the rule that the number of $d$ quarks in stable nuclei is much the same as that of $u$ quarks is violated.
The excess of $n_d$ becomes progressively larger, until the stability of nuclei ruins at $Z=82$, as shown in the Segr\`e chart, Figure~\ref{stable}, where the so-called `drip line' of stable nuclei terminates.
What is the reason for the departure from the rule (\ref{n-approx-n})? 
And then, why does Nature fail to preserve the nuclear stability above $Z=82$?

Let us return to Eq.~(\ref{R-nucleus}).
At first glance this relationship is inherent in a classical liquid drop rather than a 
quantum-mechanical system whose extension given by its Compton wavelength is inversely proportional
to ${A}$.
However, in fact it agrees well with the Pauli exclusion principle: identical fermions tend to move apart until the overlap of their wave functions becomes negligible, so that identical quarks occupy the volume proportional to their number.
Here is just what  Eq.~(\ref{R-nucleus}) manifests. 

Based on the Yukawa idea that the strong nuclear forces between nucleons have their origin in meson exchanges, it is difficult if not impossible to realize the stability of heavy nuclei.
Indeed, the Yukawa potential $-g^2e^{-m_\pi r}/r$ can only bind adjacent nucleons, and hence bulky aggregates of nucleons tend to decay under the action of the long-range inter-proton Coulomb repulsion.
Furthermore, the heavier is the nucleus, the greater is the excess of neutrons, and with it the added degeneracy pressure, causing the balance of forces problematic.
And yet stable isotopes are abundant up to $A=208$.

The conventional explanation for the end of nuclear stability at ${}^{208}_{82}{\rm Pb}$---which is that the stable balance between the Yukawa attraction and the Coulomb repulsion saturates at $A=208$---is actually difficult to regard as a serious explanation.
The strange thing about this explanation is that the saturation occurs for ${}^{208}_{82}{\rm Pb}$, having $Z/N\approx 0.65$, rather than for nuclear systems where the Coulomb repulsion is most efficient, say, for ${}^{40}_{20}{\rm Ca}$, having $Z/N=1$. 
In the long run, can this Yukawa paradigm clarify the fact that stable nuclei composed of 
only neutrons are unfeasible under terrestrial conditions?

By contrast, the existence of stable heavy nuclei is explicable in the context of an effective theory based on the action (\ref{QCD-Lagrangian}).
What counts is that widely separated quarks are subject to an infrared QCD effect steeply 
enhancing their mutual attractions.
This is the major distinction between the treatment of nuclear structure in terms of quarks and that in terms of nucleons.

We now take a closer look at this effective theory.
With the quantum minimalism in mind, we assume that the least action contribution dominates the path integral.
This is the same as saying the wave function of a single quark $\Psi$ is described by a solution to 
the Dirac equation in which $A_\alpha$ and $\Phi$ act as background fields.

We restrict our attention to spherically symmetric 
interactions, taking the Lorentz vector potential contribution to the mean field 
to be given by only $A_0(r)$. 
The arguments in support of the assumption that the interaction of the quark 
with the mean field is spherically symmetric closely resemble those in the
usual single-particle shell model of atomic nuclei \cite{KPV-2}.
We thus proceed from the Dirac Hamiltonian 
\begin{equation}
H=-i\boldsymbol{\alpha}\cdot{\nabla}+{\mathbb I}\,U_V({r})+\beta\left[m+U_S({r})\right],
\label
{Dirac-Hamiltonian}
\end{equation}
where $\boldsymbol{\alpha}$ and $\beta$ are the standard Dirac matrices, ${\mathbb I}$ an 
identity matrix, $U_V=g_VA_0$, $U_S=g_S\Phi$. 
For completing the definition of the eigenvalue problem 
\begin{equation}
H\Psi=\varepsilon\Psi\,,
\label
{Dirac-genera}
\end{equation}
two conditions are essential to add:

\noindent
(i) $U_{V}$ and $U_{S}$ are subject to the pseudospin symmetry condition \cite{Ginocchio, Liang}, 
\noindent
\begin{equation}
U_S(r)=-U_V(r)+{\cal C}\,,
\label
{pseudospin}
\end{equation}
where ${\cal C}$ is a positive constant defined for each particular kind of nuclei;

\noindent
(ii) $|U_{V}|$ and $|U_{S}|$ grow in space, {\it e.~g.}, as the Cornell potential \cite{Cornell75} 
\begin{equation}
{V_{\rm C}(r)}=
-\frac{\alpha_s}{r}+\sigma r \,.
\label
{Cornell}
\end{equation}

With (\ref{pseudospin}), the Dirac Hamiltonian (\ref{Dirac-Hamiltonian}) becomes
\begin{equation}
H=\boldsymbol{\alpha}\cdot{\bf p}+U_V({r})({\mathbb I}-\beta)+ \beta\left(m+{\cal C}\right),
\label
{Dirac-Hamiltonian-spin}
\end{equation}
implying that $m$ is shifted, 
\begin{equation}
m\to m_{\cal C}=m+{\cal C}\,.
\label
{mass_shift}
\end{equation}
This can be interpreted as a phenomenological mechanism which converts the current quark mass to the corresponding constituent quark mass.
Henceforth $m_{\cal C}$ is regarded as the constituent quark mass, and the mark ${\cal C}$ of $m_{\cal C}$ is omitted.
 
To solve (\ref{Dirac-genera}), one separates variables in the usual way \cite{Ginocchio}.
The radial part of (\ref{Dirac-genera}) is
\begin{equation}
{f'}+\frac{1+\kappa}{r}\,f-\xi g=0\,,
\label
{Dirac_radia_f}
\end{equation}
\begin{equation}
{g'}+\frac{1-\kappa}{r}\,g+\eta f=0\,,
\label
{Dirac_radia}
\end{equation}
where the prime stands for the derivative with respect to $r$,
$\kappa=\pm(j+\frac12)$ are eigenstates of the operator ${K}=-\beta\,({\bf S}\cdot{\bf L}+1)$ which 
commutes with the spherically symmetric Dirac Hamiltonian  (\ref{Dirac-Hamiltonian}), and
\begin{equation}
\xi(r)=\varepsilon+{m}+U_S(r)-U_V(r)\,,
\label
{a-df}
\end{equation}
\begin{equation}
\eta(r)=\varepsilon-{m}-U_S(r)-U_V(r)\,.
\label
{b-df}
\end{equation}
Equation (\ref{Dirac_radia_f}) is used to express $g$ in terms of $f$.
We then substitute the result into (\ref{Dirac_radia}).
If the first derivative of $f$ is eliminated from the obtained second-order differential equation, we come to the Schr\"odinger-like equation
\begin{equation}
F''+k^2F=0\,,
\label
{1D_Schroedinger}
\end{equation}                                          
\begin{equation}
k^2={\varepsilon^2-m^2}-2U(r;\varepsilon)\,.
\label
{k-df}
\end{equation}

\begin{figure}[htb]
\centerline{\includegraphics[height=\risheight]{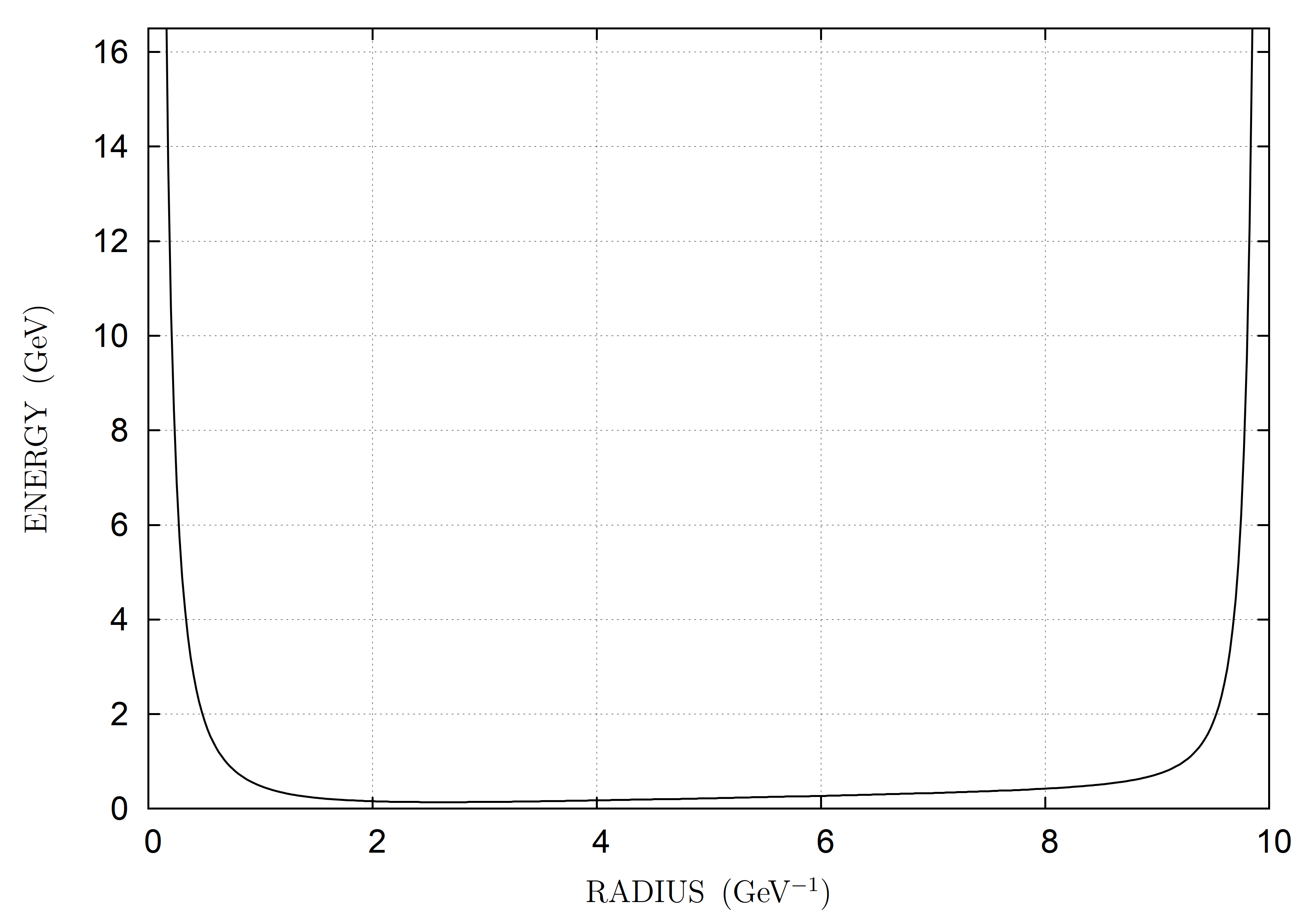}}
\caption{The effective potential  ${U}\left(r;\varepsilon\right)$ }
\label{pseudo-spin-potential}
\end{figure}

Taking $U_V=\frac12 V_{\rm C}$ and using (\ref{pseudospin}) in (\ref{a-df}) and (\ref{b-df}), we 
find the effective potential
\[U(r;\varepsilon)=\frac{1}{2{r^2}}\Biggl\{{\kappa(\kappa+1)}
+\left(\varepsilon-{m}\right)\left(-\frac{\alpha_s}{r}
+\sigma r\right){r^2}
\]
\begin{equation}
+\frac{3(\alpha_s+\sigma r^2)^2}
{4\left[\sigma r^2-\left(\varepsilon +{m}\right)r-\alpha_s\right]^2}+
\frac{\alpha_s(\kappa+1)+\kappa\sigma r^2}
{\sigma r^2-\left(\varepsilon+{m}\right)r-\alpha_s}
\Biggr\}\,.
\label
{U_eff-PSEUDO}
\end{equation}
The last two terms of (\ref{U_eff-PSEUDO}) are singular at $r=r_{\ast}$ which is
the positive root of the equation $\sigma r^2-\left(\varepsilon +{m}\right)r-\alpha_s=0$,
\begin{equation}
r_{\ast}=\frac{\left(\varepsilon +{m}\right)+
\sqrt{\left(\varepsilon +{m}\right)^2+4\sigma \alpha_s}}{2\sigma}\,.
\label
{sol}
\end{equation}
The form of ${ U}\left(r;\varepsilon\right)$ with particular values of $m$, $\varepsilon$, 
${\alpha_s}$, and $\sigma$  is depicted in 
Figure~\ref{pseudo-spin-potential}. 

From (\ref{U_eff-PSEUDO}) we notice that ${U}\left(r;\varepsilon\right)$ develops a singularity
\begin{equation}
{ U}\left(r;\varepsilon\right)\sim \gamma\left(r-r_\ast\right)^{-2},
\quad
\gamma>0\,,
\label
{effective_potential}
\end{equation}
at a finite point $r_\ast$.
Therefore, the pseudospin symmetry condition (\ref{pseudospin}) vastly enhances the interaction 
between the mean field and spin degrees of freedom of the quark to yield a cavity bounded by a 
spherical shell of radius $r_\ast$ on which ${U}\left(r_\ast;\varepsilon\right)=\infty$.
It is reasonable to identify the cavity with the interior of the nucleus, and equate $r_\ast$ to $R$ 
appearing in (\ref{R-nucleus}).
With this identification, the idea that the quark is confined to the nucleus is stated as follows: 
the probability amplitude of being somewhere in the cavity is given by solutions of the 
Schr\"odinger-like equation (\ref{1D_Schroedinger}), while that outside the cavity is 0. 
A rigorous mathematical justification of this statement is the theorem \cite{Dittrich} which asserts 
that the tunneling of a nonrelativistic particle through the one-dimensional barrier 
(\ref{effective_potential}) is forbidden iff $\gamma\ge \frac34$.
This condition is fulfilled in (\ref{U_eff-PSEUDO}).

An infinitely deep potential well similar to that shown in Figure~\ref{pseudo-spin-potential} arises whenever the mean field potential grows indefinitely with $r$ \cite{KPV-2}.
We thus take the liberty of varying the form of the mean field potential in a wide range to attain the best fit to experiment.
The Cornell potential is particularly well suited to this end.

Equation~(\ref{1D_Schroedinger}) was solved numerically using the parameters $\alpha_s=0.7$, $\sigma=0.1\,{\rm GeV}^2$, borrowed from the description of quarkonia, and $m=0.33$ {GeV}  \cite{KPV-2}.
The energy levels $\varepsilon_{n_r}$ were found, to a good approximation, to be proportional to $\sqrt{n_r}$, where $n_r$ is the radial quantum number (which is equal to the number of nodes in the radial part of the eigenfunction).
If it is granted that ${n_r}$ equals the integral part of ${A}^{2/3}$, then the cavity size $r_\ast$  scales as $\rho_0{A}^{1/3}$ with $\rho_0\approx 1$ fm.
The assumption that ${n_r}=\lbrack{A}^{2/3}\rbrack$ proves to be consistent with the basics of nuclear physics.
For instance, the comparison between the calculated magnetic moment of a single quark representing a particular nucleus and the observed magnetic moment of the nucleus itself shows the agreement within $\sim 20\%$ for a rich variety of stable isotopes \cite{KPV-2}.

{A further issue is the meaning of $\varepsilon_{n_r}$.
Although $\varepsilon_{n_r}$ stems from the eigenvalue problem for a single quark,  it seems natural to regard  this quantity as the mass of the nucleus associated with ${n_r}$.
The contribution of other quarks to $\varepsilon_{n_r}$ shows up in $U\left(r;\varepsilon_{n_r}\right)$.}

We next turn to the experimental fact that the excess of $n_d$ in stable nuclei, beginning with  
${}^{40}_{20}{\rm Ca}$, increases with $Z$.
To evaluate the infrared enhancement of the inter-quark attraction in nuclei with $Z\ge 20$ we 
consider a sequence of stable nuclei from ${}^{40}_{20}{\rm Ca}$ to ${}^{208}_{82}{\rm Pb}$.
Their mass number $A$ runs from $40$ to $208$, which corresponds to $12\le n_r\le 35$. 
Let $F_{n_r}(r)$ be the normalized solutions of (\ref{1D_Schroedinger}) for such $n_r$ 
(the explicit form of $F_{n_r}(r)$ can be found in \cite{KPV-2}).
The average energy density of the mean field 
\begin{equation}
\langle{\mathfrak{u}}\rangle
=\frac{3}{4\pi\left(\rho_0\sqrt{n_r}\right)^3}\int_0^{r_\ast}dr\,{ U}\left(r;\varepsilon_{n_r}\right)
|F_{n_r}(r)|^2
\label
{energy density}
\end{equation}
must be compared with the energy density ${\cal E}$ related to the degeneracy pressure.
To make the comparison we invoke the phenomenological fact, apparent in Figure~\ref{stable}, that the excess of neutrons ${\Delta N}$ which increases along the drip line in the range $20\le Z\le 82$ is approximately linear in $Z$: 
\begin{equation}
\Delta N=0.71\left(Z-20\right)=0.71\left(0.5 A -\Delta N -20\right). 
\label
{Delta N}
\end{equation}
Combining (\ref{Delta N}) with (\ref{deg_pressure}) gives     
\begin{equation}
{P}\propto 
\left[\left(0.29+\frac{8.3}{n_r^{3/2}}
\right)^{5/3}\frac{1}{m_p}+\left(0.71-\frac{8.3}{n_r^{3/2}}\right)^{5/3}\frac{1}{m_n}\right].
\label
{deg_pressure_to_compare}
\end{equation}                                           

\begin{figure}[htb]
\centerline{\includegraphics[height=\risheight]{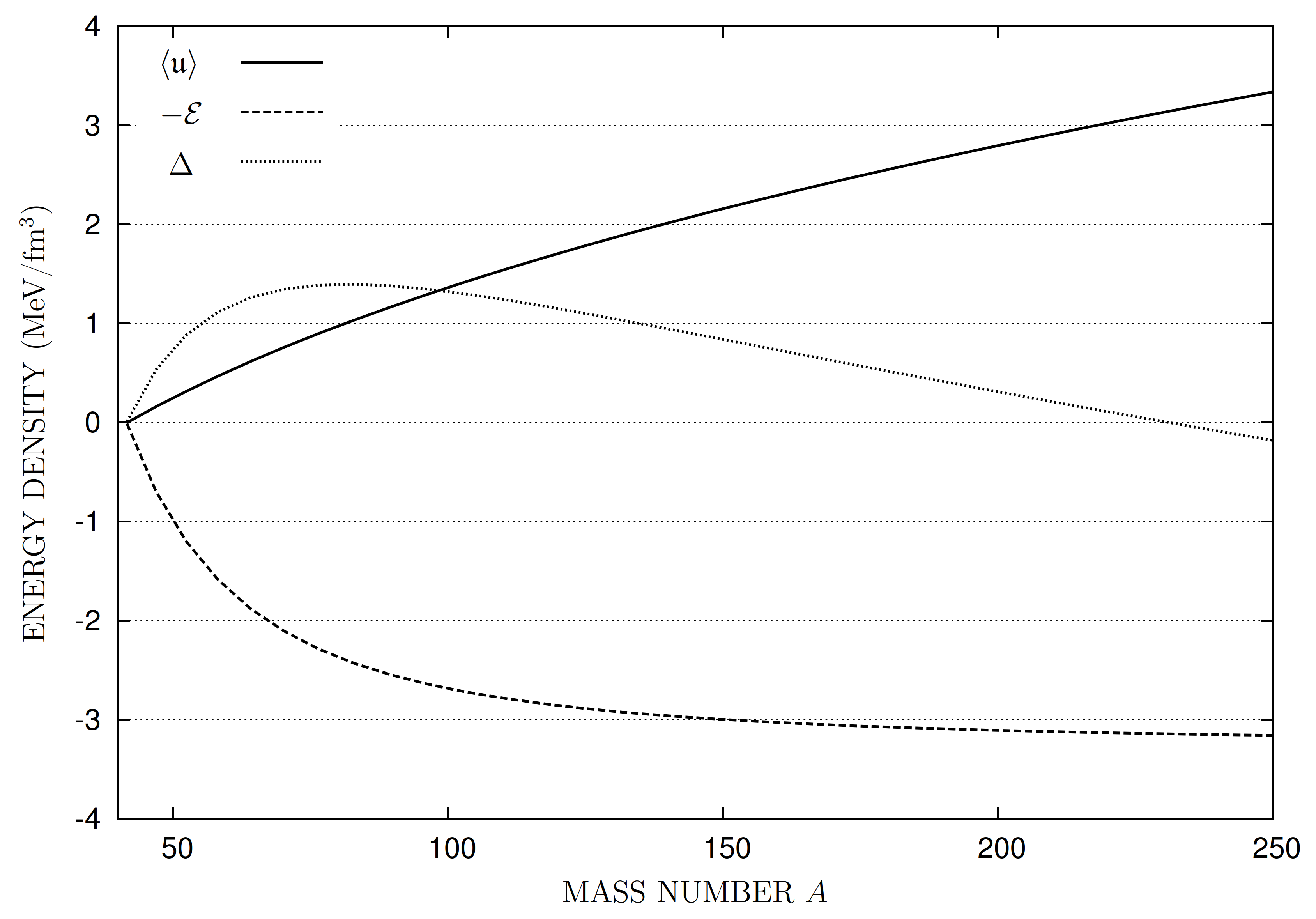}}
\caption{Average energy density of the mean field $\langle{\mathfrak{u}}\rangle$,  
energy density ${\cal E}$ related to the degeneracy pressure, 
and the difference in the magnitudes of  $\langle{\mathfrak{u}}\rangle$ and  ${\cal E}$, $\Delta=\langle{\mathfrak{u}}\rangle-{\cal E}$}
\label{comparison}
\end{figure}

The results of numerical evaluations of $\langle{\mathfrak{u}}\rangle$ and ${\cal E}=\zeta P$, together with the difference in the magnitudes of  $\langle{\mathfrak{u}}\rangle$ and  ${\cal E}$, $\Delta=\langle{\mathfrak{u}}\rangle-{\cal E}$, are plotted in Figure~\ref{comparison}. 
We take, as the starting point, the coinciding values of $\langle{\mathfrak{u}}\rangle$ and ${\cal E}$  at $A=40$, and represent only their further developments. 

\begin{figure}[htb]
\centerline{\includegraphics[height=\risheight]{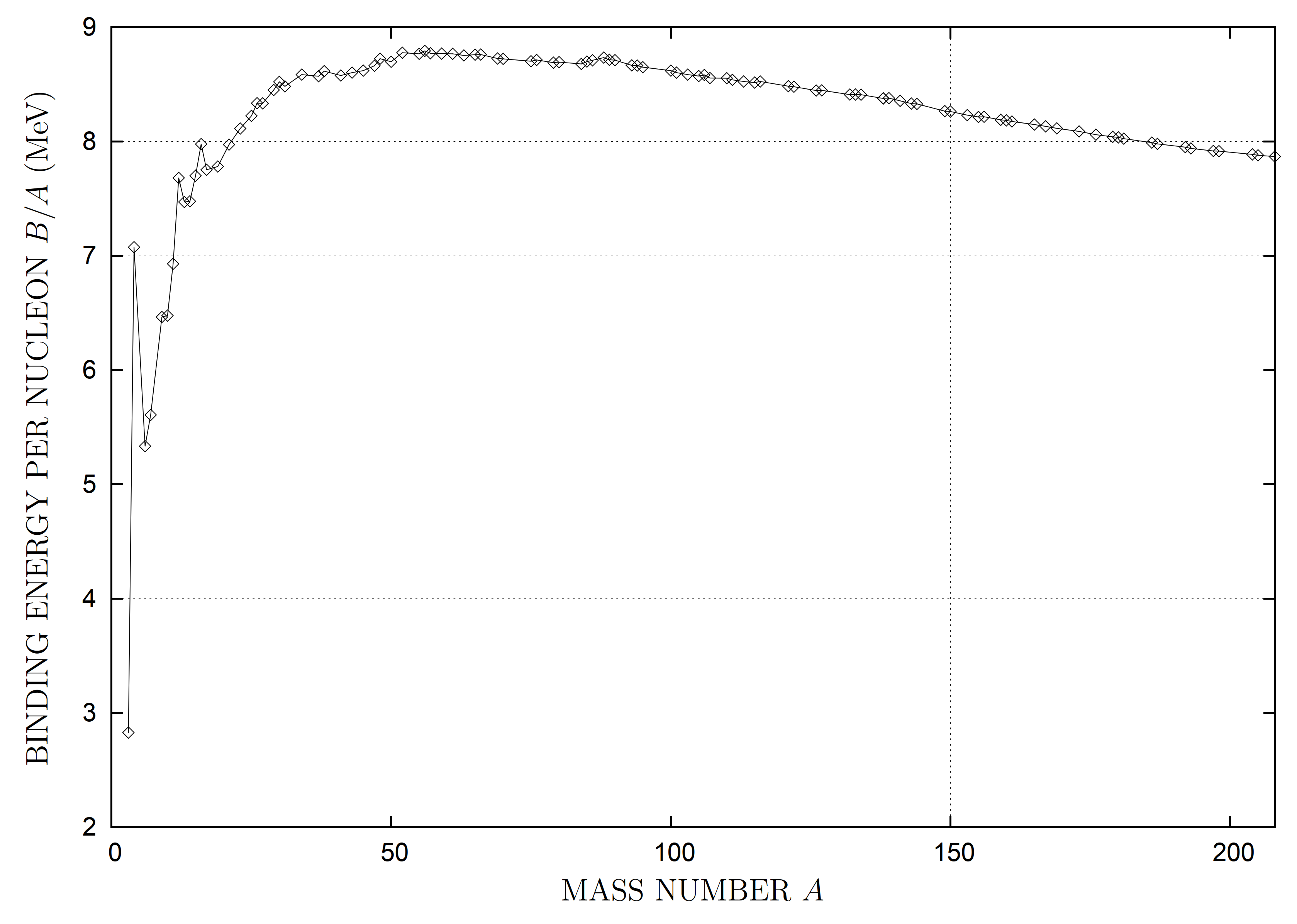}}
\caption{Average binding energy per nucleon in stable nuclei}
\label{binding}
\end{figure}

There are good grounds to believe that the infrared inter-quark attraction and degeneracy pressure play the defining roles in the force balance.
It is seen, however, that the infrared enhancement of the inter-quark attraction does not completely neutralize the increment of degeneracy pressure.
In fact, the behavior of $\Delta(A)$ has much in common with the run of the curve 
representing the average binding energy per nucleon, $B/A$, in the region 
$B/A\ge 8$ MeV~\footnote{$\Delta(A)$ peaks at $A\approx 68$, whereas the maximum of $B/A$ falls on $A\approx 62$. The difference is apt to be attributable to the neglect of the inter-quark Coulomb interactions.}, Figure~\ref{binding}.

We thus see that switching between the model assigning the responsibility for the static properties 
of a nucleus to a single quark, which resides in this nucleus and is propelled by the mean field 
generated by all other constituents of the nucleus, and the model of a Fermi gas as applied to 
quarks confined in an infinitely deep potential well of size $r_\ast$, is a self-consistent and 
phenomenologically justified procedure.

If the view of a nucleus as a collection of nucleons is abandoned, the usual concept of binding energy per nucleon loses its meaning. 
Instead, it seems reasonable to refer to $B/A$ as the deficit of the constituent quark mass in nuclei as opposed to that in free nucleons.
In going from one element of the periodic table to the next, the increase (decrease) of $B/A$ 
is suggestive of the decrease (increase) of the constituent quark mass $\sqrt{p^2}$: the greater is $B/A$, the lower is $\sqrt{p^2}$.
Recall that a classical self-interacting 
particle in the SU$({\cal N})$ Yang--Mills--Wong theory \cite{k-07}, \cite{k-19} has the four-momentum squared
\begin{equation}
p^2=m^2\left(1+\ell^2 a^2\right),
\label
{p-mu-dress-quark-sqr}
\end{equation}                                          
where $m$ is the renormalized mass (which is appropriate to define as the constituent quark mass in a free nucleon, $m\approx 330$ MeV), $a^2$  the four-acceleration squared, and $\ell$  a length characteristic of this dressed particle in the `cold phase',
\begin{equation}
\ell=\frac{8}{3m g_{\rm {YM}}^2}\left(1-\frac{1}{{\cal N}}\right),
\label
{tau-0-quark}
\end{equation}
with $g_{\rm {YM}}$ being the Yang--Mills coupling constant. 
Figure~\ref{binding} shows that the dynamics of dressed quarks in light nuclei is so much whimsical that  $\sqrt{p^2}$ may take an abrupt leap in passing to adjacent nuclides. 
However, for $A>40$, the variation of $\sqrt{p^2}$ becomes gentle, which is to say that the changes of dynamical regimes are regular.

A noticeable implication of this interpretation is that the properties of self-interacting quarks in nuclei depend on their QCD environment.
What this means is an $u$ quark in a deuteron is not identical to an $u$ quark in a lead nucleus.
Nevertheless, all $u$ quarks in a given stable nucleus are identical and indistinguishable, and the same is true of $d$ quarks. 

We are coming now to the discussion of the fact that stable elements of the periodic table are limited in number.
The mechanism of stability goes wrong in a bulky nucleus when the wave functions of the most widely separated identical quarks in this nucleus no longer overlap, which releases these quarks from control of the Pauli exclusion principle. 
Accordingly, the degeneracy pressure does not increase in proportion to the number of quarks.
To gain an impression of how the quark wave function is distributed around a stable nucleus we estimate the maximum allowable separation of quarks in stable heavy nuclei at about 7 fm by applying  Eq.~(\ref{R-nucleus}) to ${}^{208}_{82}{\rm Pb}$, and take into account that the standard deviation of this wave function for the constituent quark mass $m\approx 330$ MeV is $\approx 0.6$ fm. 
However, there are too many phenomenological aspects and empirical parameters of the developed model to mount a frontal attack on the problem of $Z_{\rm max}$, which sends us in search of more sophisticated approaches. 

\section{Holographic nuclear physics}
\label
{Holographic}
\subsection{The physics of extremal BHs maps onto that of stable nuclei}
\label
{general holographic}
The basic prescription for the holographic mapping is to identify the generating functional for 4-dimensional Green's functions in the gauge theory $W_{\rm gauge}$ with its 5-dimensional dual $Z_{\rm gravity}$ at the boundary of AdS${}_5$ \cite{Witten}, \cite{Gubser},
\begin{equation}
Z_{\rm gravity}[\chi]= W_{\rm gauge}[\Psi]\,. 
\label
{basic_prescription}
\end{equation}                                           
Here, $\chi$ is taken to be a fermionic {Dirac field}.
As for the gauge side, $\Psi$, the quark field appearing in the effective theory to low-energy QCD (with due regard for the screening effect indicated in Introduction) is the counterpart of $\chi$.

To adapt this prescription to the semiclassical, feeble-quantum dynamics that governs a Dirac particle driven by gravitational and electromagnetic fields of an extremal BH~\footnote{The regime of evolution for an extremal BH is not only semiclassical---which yet allows creations and annihilations of particles near the event horizon of an ordinary BH amenable to Hawking radiation---but also {feeble-quantum}, that is, immune of such processes.} in ${\rm AdS}_5$, we require that the {least action} contribution {dominates} the {path integral} for the generating functional, 
\begin{equation}
Z_{\rm gravity}\sim e^{-{\bar I}[\chi]}\,, 
\label
{semiclassic_prescription}
\end{equation}                                           
where ${\bar I}[\chi]$ is the Euclideanized least action of the 
Einstein--Maxwell--Chern--Simons--Dirac theory~\footnote{We take the notations and conventions that were adopted in \cite{Wu}.
The metric signature in (\ref{EMD_action}) is $(+1,-1,-1,-1,-1)$.
The curvature radius $l$ of AdS${}_5$ will be further put equal to 1.
Latin letters $A,B,\ldots$ denote local orthonormal Lorentz frame indices $0,\dots,5$, while Greek letters $\alpha,\dots$ run over five indices of spacetime  coordinates
$\{t,r,\theta,\phi,\psi\}$. 
$\chi$ is a four-component Dirac spinor, $e^\alpha_A$ a pentad, $\Gamma_\alpha$ the spinor connection.
$A_\alpha$ denotes the 5-dimensional vector potential. 
The set of matrices $\gamma^A$ is spanned by the quartet of Dirac $4\times 4$-matrices and $\gamma^5$, which realize the 5-dimensional Clifford algebra, $\{\gamma^A,\gamma^B\}=2\eta^{AB}$. 
The 5-dimensional Clifford algebra has two reducible representations, so that the Dirac field in (\ref{EMD_action}) can be treated in the 4-dimensional context, with $\gamma^5$ being the fifth basis vector component, and the spinor connection is given by a so$(1,4)$-valued 1-form $\Gamma_A=e_A^\alpha\Gamma_\alpha=\frac14\gamma^B\gamma^Cf_{BCA}$, where $f_{BCA}$ is the structure constants of so$(1,4)$.}, the simplest nontrivial descendant of the type II superstring theory,
\[
I=\int d^5x\,\Biggl\{\frac{1}{16\pi}\left[\sqrt{-g}\left(R+\frac{12}{l^2}-F^{AB}F_{AB}\right) -\frac{2}{3\sqrt{3}}\,\epsilon^{ABCDE}F_{AB} F_{CD}A_E\right]
\]
\begin{equation}
+{\chi}^\dagger\left[\gamma^A e^\alpha_A\left(\partial_\alpha+\Gamma_\alpha-iqA_\alpha\right)+\frac{i}{4\sqrt{3}}\gamma^\mu\gamma^\nu F_{\mu\nu}+\mu
\right]\chi\Biggr\}\,.
\label
{EMD_action}
\end{equation}                                           
This is the same as saying the wave function $\chi(x)$ of the Dirac particle is described by a solution to the Dirac equation
\begin{equation}
\left[\gamma^A e^\alpha_A\left(\partial_\alpha+\Gamma_\alpha-iqA_\alpha\right)+\frac{i}{4\sqrt{3}}\gamma^\mu\gamma^\nu F_{\mu\nu}+\mu\right]\chi(x)=0\,, 
\label
{Dirac_bulk}
\end{equation}                                           
where $\Gamma_\alpha$ and $A_\alpha$ represent the gravitational and electromagnetic background fields of a BH, $q$ and $\mu$ are, respectively, the charge and the mass of the Dirac particle.
The presence of Pauli term with the anomalous magnetic moment equal to $1/4\sqrt{3}$ is necessary \cite{Wu} for a separation of variables in the Dirac equation (\ref{Dirac_bulk}).

Just as the {least action} contribution, $\exp\left(-{\bar I}[\chi]\right)$, {dominates} the {path integral} for the partition function in ${\rm AdS}_5$, so does its dual in ${\mathbb R}_{1,3}$, 
\begin{equation}
W_{\rm gauge}\sim e^{-{\bar{\cal S}}[\Psi]}\,. 
\label
{semiclassic_prescription_s}
\end{equation}                                           
Here, ${\bar{\cal S}}[\Psi]$ is the Euclideanized least action (\ref{QCD-Lagrangian}).

This treatment of gauge/gravity correspondence is greatly {simplified} as against that in the general case \cite{KPV-3}.
Now, the duality implies that stationary solutions to (\ref{Dirac_bulk}) exhibit characteristic features similar to those of the corresponding solutions to (\ref{Dirac_radia_f})--(\ref{Dirac_radia}).

Our strategy is as follows.
We consider a Dirac particle moving in the gravitational and electromagnetic background 
generated by a charged, rotating, extremal BH in ${\rm AdS}_5$.
The parameters of the BH are to be such that the effective potential ${\cal U}(r)$ stemming from this background be an infinitely deep potential well similar to that depicted in Figure~\ref{pseudo-spin-potential}. 
The parameter flow may lead to a naked singularity forbidden by the weak cosmic censorship \cite{Penrose}.
If one strips the BH naked by shrinking its event horizon to a point, then the classical-quantum setup of this BH \cite{K-2008} is violated, the minimalistic quantum-mechanical status quo of the dual nucleus is disturbed, and the nuclear persistence ruins. 
This may hopefully be a suitable method for tackling the problem of $Z_{\rm max}$. 
  
We proceed from the general solution of the Einstein--Maxwell--Chern--Simons theory describing the gravitational and electromagnetic fields of a charged, rotating BH in AdS${}_5$ \cite{Cvetic}. 
The radial part of the metric, expressed in terms of the Boyer--Lindquist coordinates, is 
\begin{equation}
g_{rr}=r^{2}\left({r^2+a^2\cos^2\theta+b^2\sin^2\theta}\right){\Delta_r}^{-1}\,,
\label{metric}
\end{equation}
\begin{equation}
{\Delta_r}=\left(r^2+a^2\right)\left(r^2+b^2\right)\left({r^2}+1\right)-2Mr^2
+{(Q+ab)^2-a^2b^2}\,,
\label{Delta-r}
\end{equation}
where the parameters $M, Q, a, b$ are related to the mass, charge, and two independent angular momenta of the BH.
Our interest is with the unique positive root (two merged positive roots) of ${\Delta_r}=0$ which defines the event horizon of an extremal BH.
We thus solve the equation  
\begin{equation}
r^6+{\mathfrak a}r^4+{\mathfrak b}r^2+{\mathfrak c}=0\,,
\label
{6-order-eq}
\end{equation}
 where
\begin{equation}
{\mathfrak a}=a^2+b^2+1\,,
\label
{A-df}
\end{equation}
\begin{equation}
{\mathfrak b}=a^2+b^2-2M+a^2b^2\,,
\label
{B-df}
\end{equation}
\begin{equation}
{\mathfrak c}=\left(Q+ab\right)^2\,,
\label
{C-df}
\end{equation}
to give the unique positive solution $r=r_0$,
\begin{equation}
3r_0^2=\sqrt{{\mathfrak a}^2-3{\mathfrak b}} -{\mathfrak a}\,.
\label
{root}
\end{equation}
The condition that $r_0$ is a real double root, 
\begin{equation}
\left(2{\mathfrak a}^2-9{\mathfrak a}{\mathfrak b}+27{\mathfrak c}\right)^2
=4\left({\mathfrak a}^2-3{\mathfrak b}\right)^3,
\label
{multiple-root}
\end{equation}
combined with (\ref{root}) resuls in
\begin{equation}
Q+ab=r_0^2\sqrt{2r_0^2+{\mathfrak a}}\,.
\label
{Q}
\end{equation}

The Dirac equation (\ref{Dirac_bulk}) in the metric found in \cite{Cvetic} can be decoupled 
\cite{Wu} into temporal, radial and angular parts using the ansatz
\begin{equation}
\sqrt{r+ip\gamma^5}\chi=
e^{i\left(m\phi+k\psi-Et\right)}
\left(
\begin{array}{c}
R_2(r)S_1(p)\\
R_1(r)S_2(p)\\
R_1(r)S_1(p)\\
R_2(r)S_2(p)\\
\end{array}
\right),
\label
{psi-separ}
\end{equation}
$p=\sqrt{a^2\cos^2\theta+b^2\sin^2\theta}$, $m$ and $k$ are constants (analogous to the magnetic quantum numbers) which are associated with two independent angular momenta $a$ and $b$.
The ordinary differential equations for the radial part read
\begin{equation}
\sqrt{\Delta_r}{\cal D}_r^-R_1
=r\left\{\lambda+i\mu r-\frac{Q+ab}{2r^2}-\frac{i}{r}\left[abE-mb(1-a^2)-ka(1-b^2)\right]\right\}R_2\,,
\label
{R_1}
\end{equation}
\begin{equation}
\sqrt{\Delta_r}{\cal D}_r^+R_2
=r\left\{\lambda-i\mu r-\frac{Q+ab}{2r^2}+\frac{i}{r}\left[abE-mb(1-a^2)-ka(1-b^2)\right]\right\}R_1\,.
\label
{R_2}
\end{equation}
Here,
\begin{equation}
{\cal D}_r^\pm=\frac{d}{dr}+\frac{{\Delta'}_r}{4\Delta_r}\pm\frac{iD}{\Delta_r}\,,
\label
{D_pm}
\end{equation}
$\lambda$ is a separation constant, and
\[
D=Er^4+\left[(a^2+b^2)E-ma(1-a^2)-kb(1-b^2)-\frac{\sqrt{3}}{2}\,qQ\right]r^2
\]
\begin{equation}
+(Q+ab)\left[abE-mb(1-a^2)-ka(1-b^2)\right].
\label
{D_df}
\end{equation}

The transformation
\begin{equation}
\left(
\begin{array}{c}
f\\g
\end{array}
\right)
=
\left(
\begin{array}{cc}
1 & 1\\
i & -i\\
\end{array}
\right)
\left(
\begin{array}{c}
R_1\\
R_2
\end{array}
\right)
\label
{complex-to-real}
\end{equation}
converts (\ref{R_1}) and (\ref{R_2}) into real-valued equations.

Let $r=r_0$ be the unique positive root of the set of equations ${\Delta_r}(r)=0$, 
${\Delta'}_r(r)=0$, and $D(r)=0$.
Then the behavior of the Dirac particle in the immediate vicinity of the event horizon, inside the extremal BH, is governed by 
\begin{equation}
\frac{d}{dr}
\left(\begin{array}{c}
f \\g\end{array}\right)
=
\frac{1}{r-r_0}
\left(\begin{array}{cc}
-\dfrac12+\dfrac{A_0}{\sqrt{c_0}} & \dfrac{D_0}{{c_0}}-\dfrac{B_0}{\sqrt{c_0}}\\[3mm]
-\dfrac{D_0}{{c_0}}-\dfrac{B_0}{\sqrt{c_0}} &-\dfrac12-\dfrac{A_0}{\sqrt{c_0}}
\end{array}\right)
\left(\begin{array}{c}f \\g\end{array}\right)
+O(1)\,,
\label
{truncated}
\end{equation}
where 
\begin{equation}
A_0=\lambda r_0-\frac{Q+ab}{2r_0}=\frac{r_0}{2}\left(2\lambda-\sqrt{2r_0^2+{\mathfrak a}}\right)\,,
\label
{A-0}
\end{equation}
\begin{equation}
c_0=15r_0^4+6{\mathfrak a}r_0^2+{\mathfrak b}=4r_0^2\left(3r_0^2+{\mathfrak a}\right)\,,
\label
{c-0-}
\end{equation}
\begin{equation}
D_0=4Er_0^3+2r_0 \left[E\left({\mathfrak a}-1\right)-ma\left(1-a^2\right)-kb\left(1-b^2\right)-
\frac{\sqrt{3}}{2}qQ\right],
\label
{D-0}
\end{equation}
\begin{equation}
B_0=\mu r_0^2-abE+mb\left(1-a^2\right)+ka\left(1-b^2\right).
\label
{B-0}
\end{equation}

The next steps are similar to those described in the previous section.
One of the two equations for $f$ and $g$ is used to express $g$ in terms of  $f$, and the 
result is substituted into the other equation.
If the first derivative of $f$ is eliminated from the resulting second-order differential equation, we obtain a one-dimensional Schr\"odinger-like equation 
\begin{equation}
F''+\frac{{\cal U}_0}{\left({r}-{r}_{0}\right)^{2}}\,F=0\,,
\label
{Schr-like}
\end{equation}
where
\begin{equation}
{\cal U}_0=
\left(\frac{D_0}{c_0}\right)^2+\frac14 -\frac{A_0^2+B_0^2}{c_0}\,.
\label
{U-0}
\end{equation}

\subsection{The occurrence of a naked singularity is dual to the end of stability for nuclei with  $Z_{\max}> 82$}

The condition that the Dirac particles are unable to tunnel through the barrier at $r=r_0$ refers to ${\cal U}_0\le -\frac34$.
Because our concern is with the end of the nuclear stability, we take the upper limit of 
this inequality, ${\cal U}_0=-\frac34$.
The sole exception to the lack of tunneling is provided by the occurrence of a naked singularity, which is realized in the limit $r_0\to 0$.
It follows from (\ref{root}) and (\ref{B-df}) that the parameters of the BH in this case are 
related by
\begin{equation}
2M=a^2+b^2+a^2b^2\,,
\label
{M-to-Q-naked}
\end{equation}
and (\ref{Q}) becomes
\begin{equation}
Q=-ab\,.
\label
{Q-naked}
\end{equation}
We consider $Q$ to be dual to $Z$, which, in view of (\ref{Q-naked}), implies that $a$ is opposite 
in sign to $b$.
Since $c_0\sim 4r_0^2{\mathfrak a}$ as $r_0\to 0$, the expression (\ref{U-0}) diverges in this limit.
To avoid the divergency, we require that the divergent terms of $\left(D_0/c_0\right)^2$ and 
$B_0^2/c_0$ cancel each other, 
\begin{equation}
\left[E\left({\mathfrak a}-1\right)-ma\left(1-a^2\right)-kb\left(1-b^2\right)-\frac{\sqrt{3}}{2}qQ\right]^2
\!\!=\!{\mathfrak a}\left[abE-mb\left(1-a^2\right)-ka\left(1-b^2\right)\right]^2.
\label
{Div-cancel}
\end{equation}
The condition ${\cal U}_0=-\frac34$ is met when 
\[
16E\left[E\left({\mathfrak a}-1\right)-ma\left(1-a^2\right)-kb\left(1-b^2\right)-\frac{\sqrt{3}}{2}qQ\right]+16{\mathfrak a}^2
\]
\begin{equation}
-{\mathfrak a}\left(2\lambda-\sqrt{{\mathfrak a}}\right)^2
+8{\mu{\mathfrak a}}\left[abE-mb\left(1-a^2\right)-ka\left(1-b^2\right)\right]=0\,.
\label
{finite-part}
\end{equation}

The consistency between the dynamical affairs in the bulk and in the screen can be attained if the energy $E$ of the Dirac particle is taken to be comparable to the mass $M$ of the BH, much as the energy $\varepsilon_{n_r}$ of a single quark is likened to the mass of the nucleus involving this quark.
We just equate $E$ and $M$, 
\begin{equation}
2E=a^2+b^2+a^2b^2\,.
\label
{E-naked}
\end{equation}

To simplify matters we assume that the regime of orbiting of the Dirac particle exhibits the maximal possible symmetry, that is, $a=-b$, and $m=k$. 
Since $\mu$ is dual to the constituent quark mass of a quark belonging to a nucleus under study, we should set  $\mu=E/3Q$.
Equations (\ref{Div-cancel}), (\ref{finite-part}), and (\ref{E-naked}) become
\begin{equation}
2\left[\left({\mathfrak a}-1\right)E-\frac{\sqrt{3}}{2}qQ\right]
=\sqrt{{\mathfrak a}}\left({\mathfrak a}-1\right)E\,,
\label
{Div-cancel-s}
\end{equation}
\begin{equation}
48E\left[\left({\mathfrak a}-1\right)E-\frac{\sqrt{3}}{2}qQ\right]+48{\mathfrak a}^2- 
3{\mathfrak a}\left(2\lambda-\sqrt{{\mathfrak a}}\right)^2
-8{{\mathfrak a}}E^2=0\,,
\label
{finite-part-s}
\end{equation}
\begin{equation}
8E=\left({\mathfrak a}-1\right)\left({\mathfrak a}+3\right).
\label
{E-naked-s}
\end{equation}

Combining (\ref{Div-cancel-s}), (\ref{finite-part-s}), and  (\ref{E-naked-s}) gives 
\begin{equation}
24\,\sqrt{{\mathfrak a}}\left[\left(2\lambda-\sqrt{{\mathfrak a}}\right)^2-16{\mathfrak a}\right]=
\left({\mathfrak a}-1\right)^2\left({\mathfrak a}+3\right)^2
\left[3\left({\mathfrak a}-1\right)-\sqrt{{\mathfrak a}}\right].
\label
{A-naked-s}
\end{equation}
There is clear evidence that  $\lambda$ is a trifle over four. 
Indeed, at the semiclassical level, we have to do with the Dirac equation (\ref{Dirac_bulk}) which can be decoupled into the radial and angular parts \cite{Wu}.
The latter is invariant under ${\rm SO}(4)$ equivalent to ${\rm SO}(3)\times{\rm SO}(3)$. 
Therefore, the classical separation constant $\lambda_{\rm class}$ is given by the sum of 
eigenfunctions of the corresponding Casimir operators, $l_1(l_1+1)+l_2(l_2+1)$.
To take account of the minimal rotation effect, pertaining equally to both ${\rm SO}(3)$ rotation 
groups, we should put $l_1=l_2=1$, so that $\lambda_{\rm class}=4$.
Since the feeble-quantum correction is inversely related to  $\lambda_{\rm class}$, the final 
result ends up with $\lambda=\lambda_{\rm class}+1/\lambda_{\rm class}=4.25$\footnote{What is the 
score about the state of the whole system `BH plus Dirac particle'?
It is likely to be the ground state because the precession of the Dirac particle and  
counterrotation of the BH annihilate each other in both ${\rm SO}(3)$ sells of this geometry.
To calculate $\lambda$ exactly, it is necessary to solve the formidable problem of the motion of 
the whole system.}.     
With this value of $\lambda$, the unique positive solution of (\ref{A-naked-s}) is ${\mathfrak a}\approx 2.58$.

From (\ref{Div-cancel-s}) and (\ref{E-naked-s}), we have
\begin{equation}
qQ=\frac{1}{8\sqrt{3}}\left({\mathfrak a}-1\right)^2\left(2-\sqrt{{\mathfrak a}}\right)
\left({\mathfrak a}+3\right).
\label
{qQ}
\end{equation}
Substituting ${\mathfrak a}\approx 2.58$ in (\ref{qQ}) gives $qQ\approx 0.396$. 

We now reenter the 4-dimensional Minkowski realm ${\mathbb R}_{1,3}$ where an $u$ quark plays the role of the Dirac particle~\footnote{Out of two species of quarks,   $u$ and  $d$, we must choose $u$ because $qQ>0$, as indicated by (\ref{qQ}).}, and regain the usual natural units.
We identify the projection of $qQ$ on ${\mathbb R}_{1,3}$ with the charge of the $u$ quark 
times the charge of the nucleus involving this $u$ quark, $Z_{\max}^{\rm H}$ (H for holographic), less the charge of the quark, 
\begin{equation}
qQ=\frac23\left(Z_{\max}^{\rm H}-\frac23\right)\alpha\,,
\label
{final-Z}
\end{equation}
where $\alpha\approx 1/137$ stands for the fine structure constant, to yield
\begin{equation}
Z_{\max}^{\rm H}\approx 0.667+137\cdot1.5\cdot 0.396\approx 82\,. 
\label
{final-Z-}
\end{equation}

\section{Concluding remarks}
\label
{Conclusion}
The treatment of nuclei in terms of quarks enables us to inquire into problems that defy solutions 
when nuclei are taken as bound systems of nucleons.
Let us sum up our results.

The first issue of our concern was gaining insight into the quark content of stable light 
nuclei---from ${}^{2}_{1}{\rm H}$ to  ${}^{40}_{20}{\rm Ca}$. 
With the idea of quantum minimalism, we came to the rule $n_u\approx n_d$ underlying the stability of nuclei in this part of the periodic table.

The next point was to explain the quark content of stable nuclei in the remaining part of the periodic table where the stability condition is inconsistent with the rule $n_u\approx n_d$. 
As $A$ increases above $A=40$, the ratio $n_d/n_u$ becomes ever growing.
A plausible explanation of this fact is as follows.
In view of the phenomenological formula ${R}={R_0}\, A^{1/3}$, the volume of a nucleus is 
proportional to the number of quarks involved.
The excess of $n_d$ is required to compensate the infrared QCD enhancement of the attraction force between widely separated quarks in any roomy nucleus by strengthening their quantum repulsion, that is, by departing from the minimum of  ${\cal E}$.
We compared this infrared attraction effect with the increment of ${\cal E}$, Figure~\ref{comparison}. 
The run of their interplay $\Delta$ bears a general resemblance to that of the binding energy per 
nucleon $B/A$ in the region $B/A\ge 8$ MeV, Figure~\ref{binding}. 

This observation persuaded us to relate $B/A$ to the deficit of mass of a quark in nuclei with 
respect to its mass in free nucleons.
What this means is dressed quarks in a deuteron are not identical to dressed quarks in a lead. 
This is scarcely surprising if it is remembered that an $u$ quark in a proton is much heavier than 
that in a $\pi_0$.
However, all dressed quarks of the same species in a given nucleus are identical and 
indistinguishable.

The last issue was the finiteness of the periodic table.
To grasp the reason for instability of any nucleus heavier than ${}^{208}_{82}{\rm Pb}$, it is 
worth noting that if the wave functions of the most widely separated quarks in a given nucleus do 
not overlap, these quarks get out of control of the Pauli exclusion principle, and the degeneracy 
pressure fails to behave appropriately to maintain the balance of forces. 
Put very simply,  heavy nuclei are found to be unstable because the line of demarcation between the 
quantum and the classical is exceeded.

We specified  $Z_{\max}$ through the use of gauge/gravity duality between the dynamical affair of 
stable nuclei in ${\mathbb R}_{1,3}$ and that of extremal BHs in ${\rm AdS}_5$.
If one strips the BH naked by shrinking its event horizon to a point, the classical-quantum 
arrangement of this BH  \cite{K-2008} fails.
The occurrence of a naked singularity is the holographic counterpart of the situation that
the Pauli exclusion principle becomes inconsistent in the dual nucleus, and hence the nuclear persistence ruins.  
On this grounds we deduced the maximum allowable electic charge for heavy nuclei compatible with nuclear stability, $Z_{\max}^{\rm H}\approx 82$.

\vskip10mm
\noindent
{\Large{\bf Acknowledgements}}
\vskip5mm
\noindent
We thank the anonymous reviewer for valuable comments and recommendations.

\end{document}